\documentclass[sigconf,nonacm]{acmart}
\AtBeginDocument{%
  }

\copyrightyear{2025}
\acmYear{2025}
\setcopyright{rightsretained}
\acmConference[IMC '25] {Proceedings of the 2025 ACM Internet Measurement Conference}{October 28--31, 2025}{Madison, WI, USA.}
\acmBooktitle{Proceedings of the 2025 ACM Internet Measurement Conference (IMC '25), October 28--31, 2025,Madison, WI, USA} \acmISBN{979-8-4007-1860-1/25/10}
\acmDOI{10.1145/XXXXXX.XXXXXX}

\newcommand{\myitem}[1]{\noindent\textbf{\textit{#1}}}

\usepackage{xcolor}

\begin{document}

\title{Poster: The Internet Quality Barometer Framework}

\author{Lai Yi Ohlsen}
\affiliation{%
  \institution{Measurement Lab\\Code for Science \& Society }
  \city{Portland}
  \state{OR}
  \country{ United States}
}
\email{laiyi@measurementlab.net}

\author{Pavlos Sermpezis}
\affiliation{%
  \institution{Measurement Lab\\Code for Science \& Society }
  \city{Portland}
  \state{OR}
  \country{ United States}
}
\email{pavlos@measurementlab.net}

\author{Melissa Newcomb}
\affiliation{%
  \institution{Measurement Lab\\Code for Science \& Society }
  \city{Portland}
  \state{OR}
  \country{ United States}
}
\email{mnewcomb@measurementlab.net}

\renewcommand{\shortauthors}{Lai Yi Ohlsen, Pavlos Sermpezis, and Melissa Newcomb}

\begin{abstract}
In this paper, we introduce the Internet Quality Barometer (IQB), a framework aiming to redefine Internet quality beyond ``speed''. IQB (i) defines Internet quality in a user-centric way by considering popular use cases, (ii) maps network requirements to use cases through a set of weights and quality thresholds, and (iii) leverages publicly available Internet performance datasets, to calculate the IQB score, a composite metric that reflects the quality of Internet experience.      
\end{abstract}








\maketitle

\section{Introduction}
With the rise of real-time interactive applications like video conferencing and gaming, and the increasing size of Web pages, the demands placed on Internet connections are more complex than ever. Yet, while the Internet itself has evolved, our mainstream understanding of how to measure its quality has failed to keep pace. 
For decades, ``speed'' (typically, throughput or bandwidth) 
has been the dominant metric for assessing Internet quality
: the faster data can move, the better we expect the performance to be. However, it overlooks the growing complexity of modern Internet use.



In this paper, we propose the Internet Quality Barometer (IQB) framework. IQB leverages existing datasets~\cite{mlab-ndt7,ookla-speedtest-open-data,cloudflare-radar}, takes a holistic approach to quality beyond “speed tests” and evaluates Internet performance across various use cases (web browsing, video streaming, gaming, etc.), each with its own specific network requirements (latency, throughput, etc.).

IQB is a comprehensive framework for collecting data and calculating a composite score, the \textit{IQB Score}, which reflects the quality of Internet experience. The IQB score is inspired by other composite metrics, such as a credit score~\cite{wikipedia_credit_score} and the Nutri-Score~\cite{wikipedia_nutri_score}, which illustrate how a single score can provide a generalized or approximate assessment.



\begin{figure}
    \centering
    \includegraphics[width=1\linewidth]{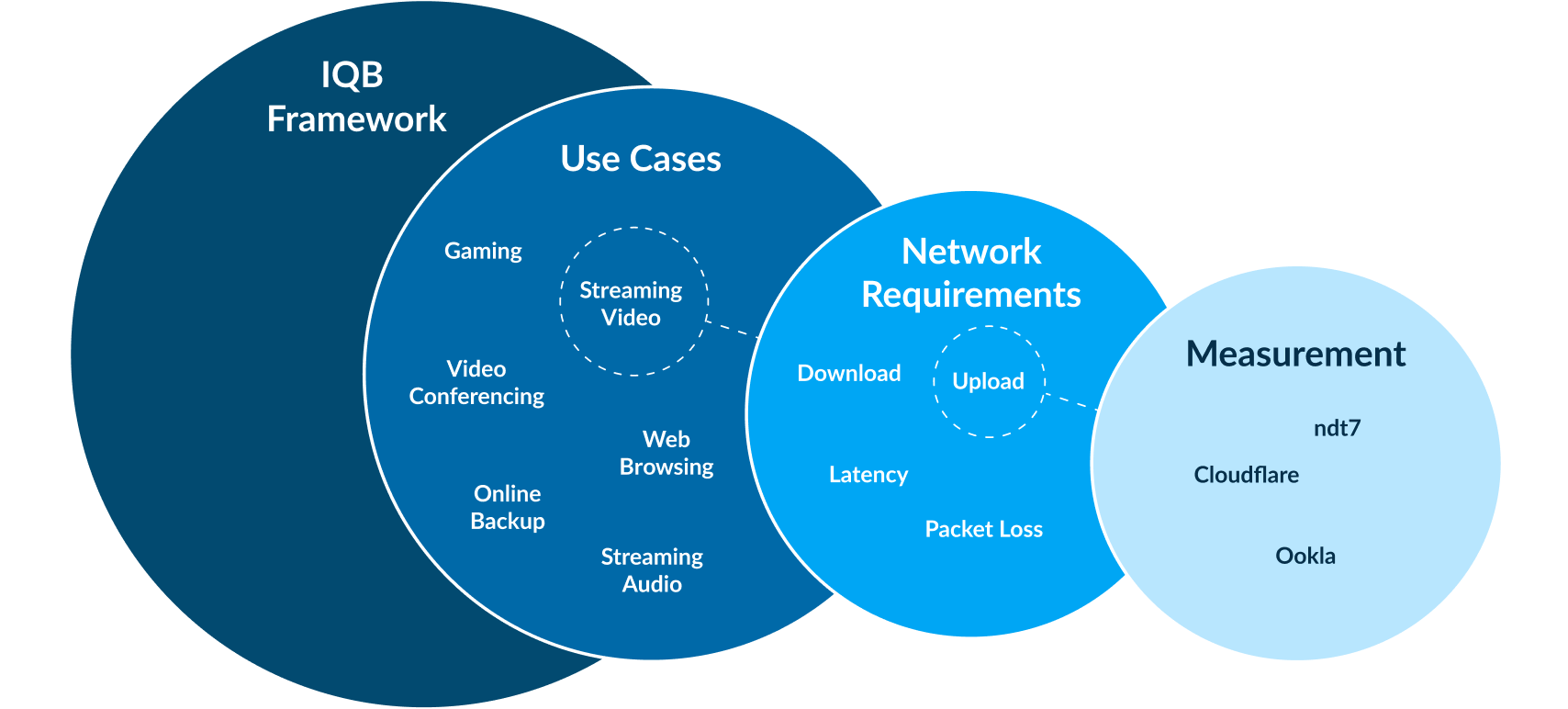}
    \caption{The IQB framework consisting of three tiers: use cases, network requirements, and datasets.}
    \label{fig:iqb-framework}
\end{figure}

\section{The IQB framework}
The IQB framework, depicted in Fig.~\ref{fig:iqb-framework}, is organized into three tiers: \textit{use cases}, \textit{network requirements}, and \textit{datasets}. 

\myitem{Use cases:} Internet users 
rarely think of Internet quality in terms of metrics like throughput, latency, or packet loss. Instead, they understand it through what the Internet enables them to do. The goal of the use cases tier is to define Internet quality in a user-centric way. Following~\cite{cranor2022making}, we consider six use cases: web browsing, streaming audio, streaming video, online backup, video conferencing, and gaming.

\myitem{Network requirements:} The goal of this layer is to map each use case into more specific technical requirements: throughput, latency, and packet loss (i.e., metrics found in openly available measurement datasets). This layer highlights nuances often overlooked in speed tests, such as the differing importance of throughput and latency depending on the use case; e.g., throughput may be critical for downloading large files, while latency is essential for video conferencing.

\begin{figure}
    \centering
    \includegraphics[width=1\linewidth]{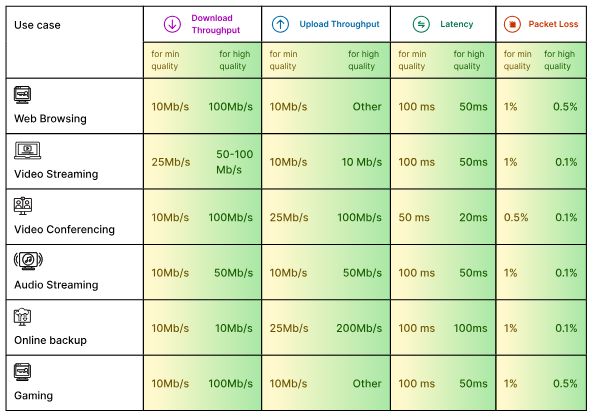}
    \caption{Network requirements thresholds for minimum and high quality for each use case}
    \label{fig:iqb-thresholds}
\end{figure}

For each use case, we provide what a user needs in terms of network requirements to be considered as having a high or minimum quality experience. To define these thresholds, we surveyed a group of experts\footnote{From Nov 2023 to Mar 2025, we engaged through interviews and workshops with more than 60 experts across various fields (network research, public policy, digital inclusion advocacy, ISPs and content providers, etc.)
.} and through their feedback arrived at the thresholds depicted in Fig~\ref{fig:iqb-thresholds}. We also asked experts to weigh in on how much a given metric matters for a specific use case, by assigning a rate of importance between 1 and 5; we present the weights in Table~\ref{table:iqb-usecase-requirements-weights}.  

\begin{table}[h!]
\centering
\small
\caption{Network requirement weights across use cases.}
\label{table:iqb-usecase-requirements-weights}
\begin{tabular}{lcccc}
\toprule
\textbf{Use Case} & \textbf{Download} & \textbf{Upload} & \textbf{Latency} & \textbf{Packet} \\
\textbf{} & \textbf{speed} & \textbf{speed} & \textbf{} & \textbf{loss} \\
\midrule
Web Browsing        & 3 & 2 & 4 & 4 \\
Video Streaming     & 4 & 2 & 4 & 4 \\
Audio Streaming     & 4 & 1 & 3 & 4 \\
Video Conferencing  & 4 & 4 & 4 & 4 \\
Online Backup       & 4 & 4 & 2 & 4 \\
Gaming              & 4 & 4 & 5 & 4 \\
\bottomrule
\end{tabular}
\end{table}

\myitem{Datasets:} Finally, we need to map network requirements to openly available datasets. Three commonly used datasets are M-Lab’s NDT~\cite{mlab-ndt7} and Cloudflare~\cite{cloudflare-radar} (available at the individual test level) and Ookla’s aggregate published data~\cite{ookla-speedtest-open-data}.

The benefit of using multiple datasets is to corroborate the insights of each other. For example, because NDT, Ookla and Cloudflare each measure throughput in a fundamentally different way, if they all signal that a connection meets the throughput requirements for gaming, then it is more likely that that connection does meet the requirements.

IQB uses the 95th percentile of a dataset to evaluate a metric: for example, to assess whether a region meets the network tier’s packet loss criteria for high-quality gaming, IQB calculates the 95th percentile of packet loss measurements collected from users in that region. The value is then compared to the predefined threshold.

\section{The IQB score}
To build an IQB score, we start at the \textit{datasets} tier, and work our way upwards through successive tiers.

\myitem{Definitions and Notations:} Let $U$, $R$ and $D$ denote the set of use cases, requirements, and datasets, respectively. 

For each dataset $d$, we aggregate its measurements (95th percentile) and calculate the \textit{binary requirement score} $S_{u,r,d}$ that indicates whether the threshold for the network requirement $r$ for a high-quality experience for use case $u$ is met.

Let $w_{u,r,d}$  denote the corresponding weight for dataset $d\in D$ under the network requirement $r\in R$ and the use-case $u\in U$. Also, let $w_{u,r}$  denote the corresponding weight for the network requirement $rR$ under the use-case $uU$, and $w_{u}$  denote the corresponding weight for the use-case $uU$ for the IQB metric. The scores are binary, i.e., $S_{u,r,d}\in\{0,1\}$, and the weights $w$ are integers between 0 and 5.
We define the normalized weights $w'\in [0,1]$ as:
\begin{equation*}
    w'_{u,r,d} = \displaystyle \frac{w_{u,r,d}}{\sum_{d} w_{u,r,d}}, \hspace{0.5cm}
    w'_{u,r} = \displaystyle \frac{w_{u,r}}{\sum_{r} w_{u,r}}, \hspace{0.5cm}
    w'_{u} = \displaystyle \frac{w_{u}}{\sum_{u} w_{u}}
\end{equation*}

\myitem{The IQB score formula:}
For each use case $u$, we calculate the \textit{requirement agreement score} per network requirement $r$, as a weighted average of the binary requirement scores:
\begin{equation}\label{eq:score-ur}
    S_{u,r} = \frac{\sum_{d}w_{u,r,d}\cdot S_{u,r,d}}{\sum_{d}w_{u,r,d}} = \sum_{d}w'_{u,r,d}\cdot S_{u,r,d} 
\end{equation}

Then we calculate the \textit{use-case score} per use case, as a weighted average of the requirement agreement scores:
\begin{equation}\label{eq:score-u-1}
    S_{u} = \frac{\sum_{r}w_{u,r}\cdot S_{u,r}}{\sum_{r}w_{u,r}} = \sum_{r}w'_{u,r}\cdot S_{u,r} 
\end{equation}
and replacing $S_{u,r}$ from equation (\ref{eq:score-ur}), we get:
\begin{equation}\label{eq:score-u}
    S_{u} = \sum_{r}w'_{u,r}\sum_{d}w'_{u,r,d}\cdot S_{u,r,d} 
    = \sum_{r}\sum_{d}w'_{u,r}\cdot w'_{u,r,d}\cdot S_{u,r,d}
\end{equation}

The IQB score is a weighted average of the use case scores
\begin{equation}\label{eq:score-iq-1}
    S_{IQB} = \frac{\sum_{w}w_{u}\cdot S_{u}}{\sum_{w}w_{u}} = \sum_{u}w'_{u}\cdot S_{u} 
\end{equation}
and replacing $S_{u}$ from equation (\ref{eq:score-u}), we get:
\begin{equation}\label{eq:score-iq}
    S_{IQB} = \sum_{u}\sum_{r}\sum_{d}w'_{u}\cdot w'_{u,r}\cdot w'_{u,r,d}\cdot S_{u,r,d}
\end{equation}

\section{Conclusion}
IQB aims to enhance our understanding of Internet quality 
and equip decision-makers with actionable insights. In this initial iteration we propose a set of choices for weights (Table~\ref{table:iqb-usecase-requirements-weights}), quality thresholds (Fig.~\ref{fig:iqb-thresholds}), and data aggregation (95th percentile), however, IQB is designed to be easily adapted (e.g, based on the intended application, or through iterative refinements based on new insights). We welcome feedback from the community for next iteratations. More details about the methodology, design choices, and insights about the IQB framework can be found in the full report~\cite{iqb-report}.

\subsection*{Acknowledgments}

IQB is funded by the Internet Society Foundation’s Research Grant program. Special thanks to John Horrigan, Amy Cesal, and Abishek Sharma. Measurement Lab is a fiscally sponsored project of the Code for Science and Society.

\bibliographystyle{ACM-Reference-Format}
\balance
\bibliography{Bibliography}

\end{document}